\documentclass[twocolumn,showpacs,preprintnumbers,amssymb,aps,pra,a4paper]{revtex4-1}
\usepackage{graphicx,rotating}
\usepackage{epsfig}
\usepackage{dcolumn}
\usepackage{bm}
\usepackage{amssymb}
\usepackage{latexsym,epsfig}
\usepackage{amsmath}
\usepackage{color}
\usepackage{amstext}

\begin{document}

\title{Relativistic coupled-cluster theory analysis of energies, hyperfine structure constants, and dipole polarizabilities of Cd$^{+}$ }
\vspace*{0.5cm}

\author{$^{a}$Cheng-Bin Li$^{\dagger}$, $^{b}$Yan-Mei Yu$^{\ddagger}$ and $^{a,c}$B. K. Sahoo}
\email{$^{\dagger}$cbli@wipm.ac.cn}
\email{$^{\ddagger}$ymyu@aphy.iphy.ac.cn}
\email{$^*$bijaya@prl.res.in}

\affiliation{$^a$State Key Laboratory of Magnetic Resonance and Atomic and Molecular Physics, Wuhan Institute of Physics and Mathematics,
Chinese Academy of Sciences, Wuhan 430071, China\\
$^b$Beijing National Laboratory for Condensed Matter Physics, Institute of Physics, Chinese Academy of Sciences, Beijing 100190,China\\
$^c$Atomic, Molecular and Optical Physics Division, Physical Research Laboratory, Navrangpura, Ahmedabad-380009, India}

\date{Received date; Accepted date}
\vskip1.0cm

\begin{abstract}
Roles of electron correlation effects in the determination of attachment energies, magnetic dipole hyperfine structure constants and electric dipole (E1) matrix
elements of the low-lying states in the singly charged cadmium ion (Cd$^+$) have been analyzed. We employ the singles and doubles approximated relativistic
coupled-cluster (RCC) method to calculate these properties. Intermediate results from the Dirac-Hartree-Fock approximation, second-order many-body
perturbation theory and considering only the linear terms of the RCC method are given to demonstrate propagation of electron correlation effects in this ion. 
Contributions from important RCC terms are also given to highlight importance of various correlation effects in the evaluation of these properties. At the end, we 
also determine E1 polarizabilities ($\alpha^{E1}$) of the ground and $5p \ ^2P_{1/2;3/2}$ states of Cd$^+$ in the {\it ab initio} approach. We estimate them again 
by replacing some of the E1 matrix elements and energies from the measurements to reduce their uncertainties so that they can be used in the high precision
experiments of this ion.

\end{abstract}
\maketitle

\section{Introduction}

 The distinct electronic structure of a singly charged cadmium ion (Cd$^{+}$) has interesting history of discovering the hollow cathode spectrum of cadmium
\cite{Cdspec1949} and preparing the Cd-vapor laser in the early days of laser physics \cite{Cdvaporlaser1970}. With the advent of ion-trapping and
laser-cooling techniques, today the singly charged ions can be well controlled and adopted for carrying out high-precision measurements \cite{KarrJPB2009,Margolis2009}.
In fact, $^{113}$Cd$^+$ ion has been now under consideration for a compact microwave atomic clock \cite{Zhang2010} for the space research and has been
achieved a fractional uncertainty of 6.6 $\times$ 10$^{-14}$ till date \cite{CdIIhysexp2015}. Also, trapping of the Cd$^+$ ion in a semi-conductor chip has
been demonstrated \cite{stick}. One of the other important applications of Cd$^+$ is, it can be used for quantum-information processing. The entanglement
between a single trapped Cd$^{+}$ with a single photon has already been realized \cite{CdIInature2004}. This ion is also observed in the interstellar
medium and the metal-poor stars by the Hubble Space Telescope \cite{ApJL1999,ApJ2012,ApJ2014}. Thus, understanding of spectroscopic properties in the Cd$^+$
ion are of immense interest.

 High-precision measurements of the ground state hyperfine structures of the $^{111}$Cd$^+$ and $^{113}$Cd$^+$ ions have been reported \cite{CdIIhysexp2006,
CdIIhysexp1996,CdIIhysexp2012,CdIIhysexp2013,CdIIhysexp2015}. Among these, measurement in the $^{113}$Cd$^+$ ion was carried out more precisely for the atomic
clock purpose \cite{CdIIhysexp2015}. Apart from the ground state, there are no precise measurements of hyperfine splitting in other excited states available.
Similarly, lifetimes of the first two excited states, 5p $^{2}P_{1/2}$ and 5p $^{2}P_{3/2}$, have been measured to quite high accuracy (about 0.4$\%$ uncertainty)
\cite{CdII5plifetime2006}. There are also measurements of some of the other excited states available with reasonable accuracies \cite{CdI-IIexpt2004}. Therefore,
it would be necessary to perform theoretical calculations of these quantities to understand the roles of electron correlation effects and predict some of the
results to guide the experimentalists for carrying out their measurements and test capabilities of many-body methods to produce these quantities.

Likewise the experiments, there are not many theoretical calculations performed to investigate the role of electron correlation effects in Cd$^+$.
Using the GRASP2K package, Yu \textit{et al.} had employed the multi-configuration Dirac-Fock (MCDF) method to calculate the low-lying resonance excitation
energies, absorption oscillator strengths and the first ionization potential of Cd$^{+}$ \cite{YuEPJD2007}. Later, Ding \textit{et al.} had extended this
calculations to give the excitation energies between some of the high-lying states and transition probabilities due to the forbidden transitions
\cite{DingJPB2012}. Also, G{\l}owacki and Migda{\l}ek had performed calculations on the oscillator strengths of the $5s_{1/2}$-$5p_{1/2,3/2}$ transitions
by employing a relativistic configuration-interaction (RCI) method \cite{GlowackiRCI2009}. Some of the excitation energies and transition matrix elements
had been reported using the third-order many-body perturbation theory (MBPT(3) method) \cite{UISafro2003} and relativistic coupled-cluster (RCC) method
with the linearized terms of singles and doubles excitations (LCCSD method) \cite{UISafro2011}. A RCC method considering important non-linear terms in the
singles and doubles excitations (CCSD method) was also applied before to give magnetic dipole hyperfine structure constants ($A_{hf}$) and lifetimes of
few states \cite{Dixithys2008}. In a recent work, third-order hyperfine induced dipole polarizabilities were estimated to determine black-body radiation
(BBR) shift for the Cd$^+$ clock transition \cite{YuBBR2017}. However, electric dipole (E1) polarizabilities ($\alpha^{E1}$) of none of the states in this 
ion are known yet. Also, the roles of electron correlation effects in the evaluation of spectroscopic properties of Cd$^+$ are not demonstrated explicitly
in any of the above theoretical studies. In this work, we intend to analyze these effects in the evaluation of energies, $A_{hf}$ values and E1 matrix elements. 
We give results considering lower-order many-body methods for better understanding of how correlation effects propagate from lower to all-order levels in 
the calculations of the above properties in the considered ion. Furthermore, we present $\alpha^{\text{E1}}$ values of the ground, 5p $^{2}P_{1/2}$ and 
5p $^{2}P_{3/2}$ states of Cd$^+$ by combining experimental energies and precise values of E1 matrix elements.

 The remaining part of the paper is organized as follows: In Sec. \ref{sec2}, we give the description of many-body methods employed in the present work. The
results are given and discussed in Sec. \ref{sec3} before concluding the work in Sec. \ref{sec4}. Unless stated otherwise, atomic units (a.u.) are used
throughout the paper.

\section{Theoretical approaches}\label{sec2}

\subsection{Computational methods}

To carry out the calculations, we consider the Dirac-Coulomb (DC) Hamiltonian which is given by
\begin{eqnarray}
 H=\sum_{i}[c\pmb{\alpha}\cdot \mathbf{p}_{i}+(\beta-1)c^2+V_{\text{nuc}}(r_{i})]+\sum_{i\le j}\frac{1}{r_{ij}} ,
\label{DCHamil}
\end{eqnarray}
where $c$ is the velocity of light, $\pmb{\alpha}$ and $\beta$ are the Dirac matrices, $V_{\text{nuc}}(r)$ is the nuclear potential, which is evaluated
adopting the Fermi charge distribution, and $\frac{1}{r_{ij}}=\frac{1}{|{\bf r}_{ij}|}=\frac{1}{|{\bf r}_i - {\bf r}_j|}$ is the two-body interaction potential between
the electrons located at ${\bf r}_i$ and ${\bf r}_j$. Correction due to the Breit interaction is estimated by adding the following potential energy
\begin{eqnarray}
V_B(r_{ij})=-\frac{\{\mbox{\boldmath$ \alpha$}_i\cdot \mbox{\boldmath$ \alpha$}_j+
(\mbox{\boldmath$ \alpha$}_i\cdot {\bf \hat{r}}_{ij})(\mbox{\boldmath$\alpha$}_j\cdot {\bf \hat{r}}_{ij}) \}}{2r_{ij}} ,
\label{VBreit}
\end{eqnarray}
where ${\bf \hat{r}}_{ij}$ is the unit vector along ${\bf r}_{ij}$. We also estimate contributions from the lower-order quantum electrodynamics (QED) interactions
due to the vacuum polarization effects and self-energy (SE) effects using the procedure described in Ref. \cite{bkscs2016}.

The ground state configuration of Cd$^+$ is [$4d^{10}$] 5$s$, which can be treated as a system with a closed-core and one valence electron outside. Many of the excited states
of this ion also have similar configurations while some of the excited states can be described as the one-hole one-particle excitation from the [$4d^{10}$] 5$s$
configuration. In this work, we would like to determine the states that are described by the closed-core with one valence electron outside. However, we take care of correlations
of electrons from these states with those are described with one-hole one-particle excitations. We discuss these frameworks in the perturbative expansion and RCC theory
below.

\begin{table*}[t]
\caption{Trends of attachment energies (in a.u.) from different approximated methods. Results including the relativistic corrections from the Breit and QED interactions
are also given. The final results from the CCSD method including higher relativistic corrections are compared with the experimental values that are quoted in the NIST database
\cite{NISTASD} and differences are given in percentage as $\Delta$.}
	\begin{ruledtabular}
		\begin{tabular}{lcccccccc}
		        transition      &   DHF          & MBPT(2)      &  LCCSD     &  CCSD          &  $+$Breit     &  $+$Breit$+$QED       &  NIST       & $\Delta$  \\
			\hline
			& & \\
                5s $^{2}S_{1/2}$& $-0.5675749$   & $-0.6259303$ & $-0.6275084$ & $-0.6197197$ & $-0.6193345$ &  $-0.6190594$ & $-0.6213690$  &  0.4\%        \\
                6s $^{2}S_{1/2}$& $-0.2326545$   & $-0.2444164$ & $-0.2445125$ & $-0.2429233$ & $-0.2428409$ &  $-0.2427841$ & $-0.2432358$  &  0.2\%        \\
                7s $^{2}S_{1/2}$& $-0.1284520$   & $-0.1329612$ & $-0.1329438$ & $-0.1323406$ & $-0.1323081$ &  $-0.1322858$ & $-0.1324703$  &  0.1\%       \\
                & & \\
                5p $^{2}P_{1/2}$& $-0.3868446$   & $-0.4203548$ & $-0.4237701$ & $-0.4184912$ & $-0.4181088$ &  $-0.4180743$ & $-0.4202703$  &  0.5\%        \\
                6p $^{2}P_{1/2}$& $-0.1816390$   & $-0.1900302$ & $-0.1905469$ & $-0.1892563$ & $-0.1891522$ &  $-0.1891432$ & $-0.1898367$  &  0.4\%       \\
                5p $^{2}P_{3/2}$& $-0.3775864$   & $-0.4087151$ & $-0.4121132$ & $-0.4070980$ & $-0.4068669$ &  $-0.4068698$ & $-0.4089593$  &  0.5\%        \\
                6p $^{2}P_{3/2}$& $-0.1787987$   & $-0.1867903$ & $-0.1873862$ & $-0.1861016$ & $-0.1860382$ &  $-0.1860394$ & $-0.1867692$  &  0.4\%       \\
                  & & \\
                5d $^{2}D_{3/2}$& $-0.2057042$   & $-0.2123090$ & $-0.2130057$ & $-0.2123385$ & $-0.2123199$ &  $-0.2123194$ & $-0.2127147$  &  0.2\%        \\
                5d $^{2}D_{5/2}$& $-0.2050792$   & $-0.2115584$ & $-0.2122487$ & $-0.2116222$ & $-0.2116250$ &  $-0.2116246$ & $-0.2120107$  &  0.2\%        \\
                  & & \\
                4f $^{2}F_{5/2}$& $-0.1254789$   & $-0.1270271$ & $-0.1273479$ & $-0.1270349$ & $-0.1270362$ &  $-0.1270350$ & $-0.1273736$  &  0.3\%       \\
                4f $^{2}F_{7/2}$& $-0.1254936$   & $-0.1270419$ & $-0.1273539$ & $-0.1270498$ & $-0.1270502$ &  $-0.1270495$ & $-0.1273145$  &  0.2\%      \\
           \end{tabular}
	\end{ruledtabular}
	\label{EnLevels}
\end{table*}

We adopt Bloch's prescription \cite{Lindgrenbook} to build-up the perturbative series of the atomic wave function starting with the mean-field wave function 
$|\Phi_v \rangle$, which is obtained by the Dirac-Hartree-Fock (DHF) approximation, by expressing as
\begin{eqnarray}
 \vert \Psi_v \rangle = \Omega_v \vert \Phi_v \rangle,
 \label{BlochPrescri}
\end{eqnarray}
where $\Omega_v$ is known as the wave operator \cite{Lindgrenbook} that is responsible to account for the residual interactions neglected in the DHF method. Since all the
interested atomic states have the same closed-core with one valence electron on different orbitals, we construct as $|\Phi_v \rangle=a_v^{\dagger} |\Phi_0 \rangle$ with
$|\Phi_0 \rangle$ representing the DHF wave function of the closed-core ($V^{N-1}$ potential). In the MBPT method, we express
\begin{eqnarray}
 \Omega_v =  \sum_k \Omega_v^{(k)},
\end{eqnarray}
where superscript $k$ denotes for the order of residual Coulomb interactions taken into account in the wave operator. Amplitudes of these wave operators are
solved successively for higher orders using the Bloch's equation, which is given by \cite{Lindgrenbook}
\begin{eqnarray}
 [\Omega_v^{(k)},H_{0}]|\Phi_v \rangle &=&  [V_{res} \Omega_v^{(k-1)}  - \sum_{m=1}^{k-1} \Omega_v^{(k-m)} E_v^{(m-1)}] |\Phi_v \rangle. \nonumber
\end{eqnarray}
In the above expression, $H_{0}$ stands for the DHF Hamiltonian,  and $E_v^{(m-1)}=\langle \Phi_v | V_{res} \Omega_v^{(m-2)}| \Phi_v\rangle$ is the $m^{th}$ order energy.

In the RCC theory framework, the wave operator follows the exponential $ansatz$ \cite{Lindgrenbook}. Therefore, we can express as
\begin{eqnarray}
 | \Psi_v \rangle  \equiv  \Omega_v \vert \Phi_v \rangle &=& e^{ \{ T_v \} } | \Phi_v \rangle ,
 \label{eqcc}
\end{eqnarray}
with the excitation operator $T_v$. Without loss of generality, we further express $T_v = T + S_v$ for computational simplicity such that $T$ represents excitations
due to correlations among the core electrons maintaining valence electron as the spectator while $S_v$ takes care of correlation of valence electron and valence
electron with the core electrons simultaneously. It, thus, yields
\begin{eqnarray}
 | \Psi_v \rangle  &=& e^{ \{ T +S_v \} } | \Phi_v \rangle \nonumber \\
                   &=& e^T \{ 1+S_v \} | \Phi_v \rangle .
\end{eqnarray}
Termination of exponential for $S_v$ series is natural here owing to presence of only one valence electron in the state. We consider only all possible singly and doubly excited
configurations through the $T$ and $S_v$ RCC operators in the (L)CCSD method approximation. This is denoted by expressing
\begin{eqnarray}
 T=T_1 +T_2 \ \ \ \text{and} \ \ \ S_v = S_{1v} + S_{2v}
 \label{eqsd}
\end{eqnarray}
in our calculations. As we had mentioned before, both the $T_1$ and $S_{2v}$ operators can generate one-hole one-particle excitations independently acting upon 
$| \Phi_v \rangle$ in these formulations. For example, when these operators act on the $[4d^{10}]5s$ and $[4d^{10}]5p_{1/2;3/2}$ configurations it can also generate
the $4d^9 5s^2$ and $4d^9 5s 5p$ configurations, respectively, along with other singles and doubles configurations. Therefore, even though we do not calculate the low-lying 
states with the $4d^9 5s^2$ and $4d^9 5s 5p$ configurations explicitly using our RCC method but their correlation contributions are taken into account implicitly while 
determining different atomic states.

The amplitudes of both the $T$ and $S_v$ operators are obtained by solving the following equations
\begin{eqnarray}
 \langle \Phi_0^* \vert \overline{H}   \vert \Phi_0 \rangle &=& 0
\label{eqt}
 \end{eqnarray}
and
\begin{eqnarray}
 \langle \Phi_v^* \vert ( \overline{H} - \Delta E_v  ) S_v \vert \Phi_v \rangle &=&  - \langle \Phi_v^* \vert  \overline{H}  \vert \Phi_v \rangle ,
\label{eqsv}
 \end{eqnarray}
where $\vert \Phi_0^* \rangle$ and $\vert \Phi_v^* \rangle$ are the excited state configurations with respect to the DHF wave functions $\vert \Phi_0 \rangle$ and
$\vert \Phi_v \rangle$, respectively, and $\overline{H}= ( H e^T )_l$ with subscript $l$ representing for the linked terms only. The attachment energy
$\Delta E_v=E_v-E_0$, for the energy $E_0$ of the $[4d^{10}]$ configuration, is evaluated by
\begin{eqnarray}
 \Delta E_v  = \langle \Phi_v \vert \overline{H} \left \{ 1+S_v \right \} \vert \Phi_v \rangle.
 \label{eqeng}
\end{eqnarray}
For the CCSD method, we consider all possible singles and doubles configurations for the excited determinants $\vert \Phi_0^* \rangle$ and $\vert \Phi_v^* \rangle$. Again,
we use the approximation $\overline{H}=H+HT+HS_v$ in order to obtain results with the LCCSD method approximation.

\begin{table*}[t]
	\caption{Calculated $A_{hf}/g_I$ values (in MHz) from different methods. Multiplying $g_I=-1.1897722$ and $g_I=-1.2446018$ in the total values from the CCSD method and
	higher relativistic corrections, the $A_{hf}$ values for the respective $^{111}$Cd$^+$ and $^{113}$Cd$^+$ ions are evaluated. These values are compared with the available
	experimental results and other calculations.}
	\begin{ruledtabular}
		\begin{tabular}{lcccccc|cc|ccc}
			states           & DHF      & MBPT(2)    & LCCSD    & CCSD    & $\Delta$Breit   &  $\Delta$QED    & \multicolumn{2}{c|}{$A_{hf}^{ ^{111}\text{Cd}^+}$} &  \multicolumn{3}{c}{$A_{hf}^{ ^{113}\text{Cd}^+}$}   \\
			\cline{8-9} \cline{10-12}\\
			                 &          &            &          &         &                 &                 &     This work & Experiment & This work & Others  & Experiment \\
            \hline
                                         &          &            &          &         &                 &                 &               &             &          &        & \\
			5s $^{2}S_{1/2}$ & 9585.78 & 12318.07 & 12821.53 & 12262.32  & $-1.80$   & $-91.42$  &  14478(175)    & 14530.507 \cite{CdIIhysexp2012}   & 15146(183)   & 15280 \cite{Dixithys2008}  & 15199.863 \cite{CdIIhysexp2015}        \\
			6s $^{2}S_{1/2}$ & 2179.67 &  2570.42 &  2668.03 &  2568.77  &  0.55   & $-17.76$  &   3036(41)     &    & 3176(43)     & 3230 \cite{Dixithys2008} &          \\
			7s $^{2}S_{1/2}$ &  877.01 &  1011.14 &  1046.84 &  1010.28  &  0.72   &  $-6.88$  &   1195(15)     &    & 1250(15)               \\
	                                 &          &            &          &         &                 &                 &               &             &          &        & \\		
			5p $^{2}P_{1/2}$ & 1477.19 &  1956.30 &  2084.25 &  1969.81  & $-6.04$   &  $-2.51$  &   2333(31)     &    & 2441(33)    & 2430 \cite{Dixithys2008} &            \\
			6p $^{2}P_{1/2}$ &  440.12 &   531.94 &   552.47 &   534.74  & $-1.02$   & $-0.61$  &    634(10)     &    & 664(10)      & 667.81   \cite{Dixithys2008} &        \\
			5p $^{2}P_{3/2}$ &  229.08 &   337.02 &   364.65 &   333.89  &  0.16   &  $-0.20$  &    397(6)      &    & 416(6)     & 406.02 \cite{Dixithys2008} & 400 \cite{CdIIhysexp1996} \\
			6p $^{2}P_{3/2}$ &   69.97 &    96.84 &   109.17 &    94.47  &  0.08   &  $-0.07$  &    112(1)      &    & 118(1)     &  118.12 \cite{Dixithys2008}         \\
			                 &          &            &          &         &                 &                 &               &             &          &        & \\
			5d $^{2}D_{3/2}$ &   28.78 &    44.18 &    49.75 &    52.30  &  0.13   &  $-0.02$  &     62.4(1.1)  &    &  65.2(1.2)            \\
			5d $^{2}D_{5/2}$ &   12.22 &    17.89 &    20.79 &    20.39  &  0.07   &   0.01  &     24.4(5)    &    &  25.5(5)              \\
                                         &          &            &          &         &                 &                 &               &             &          &        & \\
			4f $^{2}F_{5/2}$ &    0.44 &     0.25 &    $-0.08$ &     0.40  & $-0.001$  &  $-0.002$ &     0.473(12)  &    &   0.494(12)              \\
			4f $^{2}F_{7/2}$ &    0.24 &     0.06 &     0.05 &     0.05  & $-0.0014$ &   0.002 &     0.060(4)   &    &   0.063(5)              \\
        \end{tabular}
	\end{ruledtabular}
	\label{Abhfth}
\end{table*}

With the knowledge of amplitudes of the wave operators in the MBPT and RCC methods, we evaluate the transition matrix element of a general operator $O$ between the states
$\vert \Psi_i \rangle $ and $\vert \Psi_f \rangle $ using the expression
\begin{eqnarray}
\frac{\langle \Psi_f \vert O \vert \Psi_i \rangle}{ \sqrt{\langle \Psi_f \vert \Psi_f \rangle \langle \Psi_i \vert \Psi_i \rangle}}
= \frac {\langle \Phi_f \vert \Omega_f^{\dagger} O \Omega_i \vert \Phi_i\rangle}
{\sqrt{\langle \Phi_f \vert \Omega_f^{\dagger} \Omega_f \vert \Phi_f \rangle \langle \Phi_i \vert \Omega_i^{\dagger} \Omega_i \vert \Phi_i \rangle}}  . \ \ \ \ \
\label{preq}
\end{eqnarray}
The expectation values are determined by considering $\vert \Psi_i \rangle =\vert \Psi_f \rangle $ in this expression. It can also be noticed that the $\Omega_f^{\dagger} O \Omega_i$
and $\Omega_f^{\dagger} \Omega_i$ terms contain non-terminating series $e^{T^{\dagger}} O e^T$ and $e^{T^{\dagger}} e^T$ in the CCSD method. These terms are evaluated
self-consistently to infinity number as discussed in our recent works \cite{sahoo2,sahoo4}.

To estimate typical contributions from the neglected higher order excitations, we define a triple excitation RCC operator perturbatively involving the valence orbital as
\begin{eqnarray}
 S_{3v}^{pert} &=& \frac{1}{4} \sum_{ab,pqr} \frac{\big ( H T_2 + H S_{2v} \big )_{abv}^{pqr}}{\Delta E_v + \epsilon_a + \epsilon_b - \epsilon_p -\epsilon_q - \epsilon_r} ,
\label{s3eq}
 \end{eqnarray}
where ${a,b}$ and ${p,q,r}$ represent for the occupied and virtual orbitals, respectively, and the $\epsilon$s are their corresponding orbital energies. Since it involves
the valence orbital, it will give the dominant triples contributions that are neglected in the CCSD method. We include this operator in the property evaluating expression
to estimate uncertainties due to the neglected higher level excitations.

\subsection{Atomic properties concerned in this work}

We are interested to analyze the correlation trends in the hyperfine structure constants ($A_{hf}$) and E1 transition matrix elements of the low-lying states of Cd$^+$.
Using the accurate values of the E1 matrix elements, we also determine $\alpha^{E1}$ values. General expressions used for these calculations are given below.

The expression for the magnetic dipole hyperfine structure constant is given by \cite{schwartz}
\begin{eqnarray}
 A_{hf}=\mu_{N} g_{I} \frac{\langle J||{\bf T}_{hf}^{(M1)}||J\rangle}{\sqrt{J(J+1)(2J+1)}},
 \label{Ahf}
\end{eqnarray}
where $\mu_{N}$ is the nuclear magneton and $g_{I}$ is the ratio of nuclear magnetic dipole moment $\mu_{I}$ and the nuclear spin $I$.
The single particle matrix element of the hyperfine interaction operator $T_{hf}^{(M1)}=\sum_i t_{hf}^{(1)}(r_i)$ is given by
\begin{eqnarray}
\langle \kappa_{f}||t_{hf}^{(1)}||\kappa_{i}\rangle &=&-(\kappa_{f}+\kappa_{i})\langle -\kappa_{f}||{\bf C}^{(1)}||\kappa_{i}\rangle \nonumber \\
&& \times \int^{\infty}_{0}dr \frac{(P_{f}Q_{i}+Q_{f}P_{i})}{r^{2}} .
\label{om1hyp}
\end{eqnarray}
We have used $g_I=-1.1897722$ and $g_I=-1.2446018$ \cite{StoneTable} for the $^{111}$Cd$^+$ and $^{113}$Cd$^+$ ions, respectively.

\begin{table*}[t]
	\caption{Contributions from different RCC terms to the $A_{hf}/g_I$ calculations (in MHz) without accounting for normalization of wave functions are given in the
	CCSD method approximation. Corrections due to the normalization of wave functions are listed under ``Norm''. Contributions from the non-linear terms that are not
	given explicitly here are added together and quoted as ``Others''. Contributions from hermitian conjugate ($h.c.$) terms are added up.}
	\begin{ruledtabular}
		\begin{tabular}{lcccccccc}
			States           & $\overline{O}$     & $OS_{1v}+h.c.$  & $OS_{2v}+h.c.$ & $S_{1v}^{\dag}OS_{1v}$ & $S_{1v}^{\dag}OS_{2v}+h.c.$  & $S_{2v}^{\dag}OS_{2v}$ & Norm   & Others   \\
            \hline
            & & \\
     			5s $^{2}S_{1/2}$ & 9539.29        &     1868.98   &  793.48         &      91.10             &   63.00                     & 258.91                 & $-295.38$ &  $-57.06$  \\
			6s $^{2}S_{1/2}$ & 2168.24        &      207.64   &  164.66         &       4.95             &    3.02                     &  63.20                 &  $-33.15$ &   $-9.79$   \\
			7s $^{2}S_{1/2}$ &  872.61        &       60.04   &   64.38         &       1.03             &    0.20                     &  27.06                 &  $-11.49$ &   $-3.55$   \\
			&& \\
			5p $^{2}P_{1/2}$ & 1466.06        &      397.86   &   94.04         &      26.88             &   13.34                     &  30.77                 &  $-42.15$ &  $-16.99$   \\
			6p $^{2}P_{1/2}$ &  437.19        &       64.16   &   31.46         &       2.38             &    2.30                     &   7.31                 &   $-7.97$ &   $-2.09$   \\
			5p $^{2}P_{3/2}$ &  229.57        &       61.58   &   26.54         &       4.16             &    3.54                     &  16.83                 &   $-6.91$ &   $-1.42$  \\
			6p $^{2}P_{3/2}$ &   70.12        &       10.48   &    8.62         &       0.40             &    0.54                     &   7.48                 &   $-1.39$ &   $-1.78$  \\
			&& \\
			5d $^{2}D_{3/2}$ &   30.20        &        7.82   &    8.84         &       0.55             &    0.80                     &   4.05                 &   $-0.41$ &   0.45    \\
			5d $^{2}D_{5/2}$ &   12.77        &        3.32   &    3.04         &       0.23             &    0.22                     &   0.65                 &   $-0.17$ &   0.33    \\
			&& \\
			4f $^{2}F_{5/2}$ &    0.44        &        0.04   &   $-0.26$       &      $\sim$ 0.00       &   $-0.04$                   &   0.30                 & $\sim$ 0.00 & $-0.08$   \\
			4f $^{2}F_{7/2}$ &    0.25        &        0.02   &   $-0.24$       &      $\sim$ 0.00       &   $-0.04$                   &   0.03                 & $\sim$ 0.00 &  0.03   \\
           \end{tabular}
	\end{ruledtabular}
	\label{ContriAhfRCCSD}
\end{table*}

We also extract E1 matrix elements (in a.u.) from the experimentally known transition probabilities (in $s^{-1}$) using the following relation
\begin{eqnarray}
 A_{if}^{E1} = \frac{2.02613 \times 10^{18}}{\lambda_{if}^{3} g_{i} } \vert \langle J_i ||{\bf D} || J_f \rangle \vert ^{2} ,
\label{E1TranProb}
\end{eqnarray}
where $g_i = 2J_i + 1$ is the degeneracy factor of the state $\vert \Psi_i \rangle $ with the angular momentum $J_i$ and $\lambda_{if}$ is the transition wavelength
in {\AA}. These values are used to compare with our calculations and also to determine $\alpha^{E1}$ values more precisely.

The expression for the static dipole polarizability is conveniently given by
\begin{eqnarray}
\alpha_i^{E1}=\alpha^{S}_{i} +\frac{3M^2_{i}-J_{i}(J_{i}+1)}{J_{i}(2J_{i}-1)}\alpha^{T}_{i},
\end{eqnarray}
where $\alpha^{S}_{i}$ and $\alpha^{T}_{i}$ are known as the scalar and tensor components of the electric dipole polarizability for the state $\vert \Psi_i \rangle $ with angular momentum
$J_i$ and its component $M_i$.

We employ the CCSD method in the equation-of-motion framework \cite{kallay} using Dyall's relativistic triple-$\zeta$ basis function \cite{Basis} from the DIRAC package
\cite{Dirac} to obtain the {\it ab initio} values of scalar and tensor polarizabilities. A finite-field approach is adopted to express the energy of the 
$| \gamma_i, J_i, M_i \rangle$ state, for the additional quantum number $\gamma_i$, in the presence of an isotropic electric field with strength in the z-direction ${\cal E}_z$ as
\begin{eqnarray}\label{polar}
E_{\gamma_i, J_i, M_i}({\cal E}_z) = E_{\gamma_i, J_i, M_i}(0) - \frac{{\cal E}_z^2}{2} \alpha_{zz}^{E1}(\gamma_i,J_i,M_i)  - \dots, \ \ \
\end{eqnarray}
where $E_{\gamma_i, J_i, M_i}({\cal E}_z)$ and $E_{\gamma_i, J_i, M_i}(0)$ are the total energies of the state in the absence and presence of the field, respectively.
Here, $\alpha_{zz}^{E1}(\gamma_i,J_i,M_i)$ is its z-component of $\alpha_i^{E1}$ and is evaluated as the second derivative of $E_{\gamma_i, J_i, M_i}({\cal E}_z)$
with respect to ${\cal E}_z$. After obtaining $\alpha_{zz}^{E1}(\gamma_i,J_i,M_i)$ values, we determine scalar polarizability by using the relation $\alpha_i^S=
\Sigma_{M_i}\alpha_{zz}^{E1}(\gamma_i, J_i, M_i)/(2J_i+1)$ and tensor polarizability as $\alpha_i^T=\alpha^{E1}_{zz}(\gamma_i,J_i,J_i)-\alpha_i^S$.

We also rewrite expressions for $\alpha_i^S$ and $\alpha_i^T$ as
\begin{eqnarray}
 \alpha_i^{S/T} = 2 \langle \Psi_i^{(0)} | \tilde{D} | \Psi_i^{(1)} \rangle ,
\end{eqnarray}
with the unperturbed wave function $|\Psi_i^{(0)} \rangle$ and its first order correction due to dipole operator $| \Psi_i^{(1)} \rangle$, to determine dipole polarizabilities 
in the spherical coordinate system using the atomic orbitals with definite parities. In this expression, we define
the respective effective dipole operator for the scalar and tensor components as described in Ref. \cite{bks-polz} to obtain the corresponding expressions for $\alpha_i^S$ and $\alpha_i^T$. Using the
prescribed formalisms in Refs. \cite{bks-polz,bks-pnc}, we solve the wave function $| \Psi_i^{(1)} \rangle$ at the DHF and MBPT(3) method
to estimate the $\alpha_i^S$ and $\alpha_i^T$ values and then compare them with the results obtained using the CCSD method to understand propagation of
electron correlation effects from lower-orders to all-orders in these calculations.

\begin{table*}[t]
	\caption{Reduced E1 matrix elements (in a.u.) of some of the important transitions in Cd$^+$ from different many-body methods. Our results are also compared
with the other reported calculations employing the DHF, MBPT(2) and LCCSD methods in Ref. \cite{UISafro2011} and values that are extracted from the lifetime
measurements \cite{CdII5plifetime2006,CdI-IIexpt2004}.}
	\begin{ruledtabular}
		\begin{tabular}{lcccccccc}
			$|\Psi_i \rangle \rightarrow |\Psi_f\rangle$      & DHF       & MBPT(2)   & LCCSD    & CCSD   &  \multicolumn{3}{c}{Ref. \cite{UISafro2011}}  & Experiment \\
			                                  \cline{6-8} \\
			                                  &           &           &          &               &    DHF &    MBPT(2)   & LCCSD$^*$                  &    \\
			                                    \hline
			                                    & & \\
			5s $^{2}S_{1/2}$ - 5p $^{2}P_{1/2}$ & 2.427   &  2.032   & 1.888    & 1.970(8)  &  2.4271 & 2.0342 & 1.9392  & 1.910(4)\cite{CdII5plifetime2006}, 1.89(3)\cite{CdI-IIexpt2004}      \\
			5s $^{2}S_{1/2}$ - 6p $^{2}P_{1/2}$ & $-0.063$   &  0.084   & 0.118    & 0.079(4)  &      &                                      &       \\
			5s $^{2}S_{1/2}$ - 5p $^{2}P_{3/2}$ & 3.428   &  2.881   & 2.678    & 2.795(11)  &  3.4280 & 2.8889 & 2.7513  & 2.713(5)\cite{CdII5plifetime2006}, 2.79(4)\cite{CdI-IIexpt2004}       \\
            5s $^{2}S_{1/2}$ - 6p $^{2}P_{3/2}$ & $-0.176$   &  0.030   & 0.087    & 0.029(3)  &           &      &                & \\
            & & \\
		    5p $^{2}P_{1/2}$ - 6s $^{2}S_{1/2}$ & 1.771   &  1.689   & 1.625    & 1.647(6)  &          &       &                   &  1.72(12)\cite{CdI-IIexpt2004}  \\	
		    5p $^{2}P_{1/2}$ - 5d $^{2}D_{3/2}$ & 4.011   &  3.521   & 3.332    & 3.475(12)  &  4.0144 & 3.7414  & 3.4401  &  3.08(13)\cite{CdI-IIexpt2004}   \\
		  & &  \\
            5p $^{2}P_{3/2}$ - 6s $^{2}S_{1/2}$ & 2.701   &  2.580   & 2.482    & 2.517(10)  &         &        &              & 2.31(9)\cite{CdI-IIexpt2004}  \\
		    5d $^{2}P_{3/2}$ - 5d $^{2}D_{3/2}$ & 1.867   &  1.649   & 1.565    & 1.628(6)  &  1.8684 & 1.7444 & 1.6122  & 1.57(6)\cite{CdI-IIexpt2004}    \\
		    5d $^{2}P_{3/2}$ - 5d $^{2}D_{5/2}$ & 5.581   &  4.928   & 4.680    & 4.868(22)  & 5.5857 &  5.2181 & 4.8195  & 4.62(5)\cite{CdI-IIexpt2004}     \\	
		\end{tabular}
	\end{ruledtabular}
	$^*$Denoted as SD method in the original paper.
	\label{E1-MEs}
\end{table*}

Our ultimate intention is to provide very precise values of polarizabilities for the experimental use. Thus, we would like to reduce the uncertainties in the {\it ab initio}
results by substituting precisely known energies and E1 matrix elements from the experimental observations. For this purpose, we also express formulas for $\alpha_i^S$ and
$\alpha_i^T$ in the sum-over-states approach as
\begin{eqnarray}
\alpha^S_{i}= \frac{2}{3(2J_i+1)} \sum_{n \ne i} \frac{ |\langle J_n || {\bf D} || J_i \rangle|^2}{E_n - E_i},
\end{eqnarray}
and
\begin{eqnarray}
\alpha^T_{i}&=& 4\left(\frac{5J_{i}(2J_{i}-1)}{6(J_{i}+1)(2J_{i}+1)(2J_{i}+3)}\right)^{1/2} \nonumber \\
&& \times \sum_n (-1)^{J_i+J_n} \left\{\begin{array}{ccc}J_i & 1 & J_n \\ 1 & J_i & 2\\ \end{array} \right\} \frac{|\langle J_n||{\bf D} || J_i \rangle|^2} {E_n - E_i}. \ \ \
\end{eqnarray}
It to be noted that for $J_i\le 1/2$ the $\alpha^{T}_{i}$ component does not contribute to $\alpha^{E1}$ owing to the properties of the above 6j symbol.

Since we are dealing with atomic states that are expressed as Slater determinants, the $|\langle J_n || {\bf D} || J_i \rangle|^2$ values will have contributions from the
core orbitals and continuum. To account for these contributions, we divide contributions to scalar and tensor components as
\begin{eqnarray}
\alpha^{S,T}_{i}=  \alpha^{S,T}_{i} (c) + \alpha^{S,T}_{i}(cv) + \alpha^{S,T}_{i}(v)
\end{eqnarray}
following the discussions in Ref. \cite{bindiya}, where the notations $c$, $cv$ and $v$ in the parentheses represent for the contributions from the closed-core, core-valence
interactions and valence correlations respectively. It can be shown that due to the presence of the phase factor $(-1)^{J_i+J_n}$ in the tensor component of the
polarizability, the closed-core contribution becomes zero. Again, $\alpha^{S,T}_{i}(v)$ will have contributions from both the bound states and continuum. The contributions
from the bound states are referred to as ``Main'' contributions while from the continuum we denote as ``Tail'' contributions.

\section{Results and Discussion}\label{sec3}

We present attachment energies of the $(5-7)S$, $(5-6)P$, $5D$ and $4F$ states of Cd$^+$ in Table \ref{EnLevels} from different methods using the DC Hamiltonian. As can be seen
the results from the CCSD method are about 0.5\% accurate compared with the experimental values, which are also quoted in the same table from the National Institute of
Science and Technology (NIST) database \cite{NISTASD}. We have also given corrections from the Breit and QED interactions estimated using the CCSD method. As can be seen,
the DHF values are about 10\% smaller than the experimental values for the $S$ and $P$ states. Inclusion of the correlation effects through the second-order MBPT (MBPT(2))
and LCCSD methods
over estimate the results compared to the experimental values while the CCSD method gives values close to the experimental results. The higher relativistic corrections to the
energies are found to be very small. Agreement between our CCSD results with the experimental values for the energies suggest that we can also obtain the hyperfine
structure constants and E1 matrix elements of the above states reliably by employing this method.

\begin{table*}[t]
	\caption{Individual contributions from different CCSD terms to the reduced E1 matrix elements (in a.u.) of the transitions given in Table \ref{E1-MEs}.}
	\begin{ruledtabular}
		\begin{tabular}{lccccccccccc}
		Transition                  & $\overline{O}$ & $OS_{1i}$ & $S_{1f}^{\dag}O$ & $OS_{2i}$&$S_{2f}^{\dag}O$ &$S_{1f}^{\dag}OS_{1i}$ &$S_{1f}^{\dag}OS_{2i}$&$S_{2f}^{\dag}OS_{1i}$&$S_{2f}^{\dag}OS_{2i}$&Norm  & Others   \\
			\hline
            & & \\
5s $^{2}S_{1/2}$ - 5p $^{2}P_{1/2}$ &  2.424         & $-0.118$  &  $-0.002$        & $-0.155$ & $-0.208$        &   0.014               &   $-0.007$           &   $-0.001$           & 0.029                &$-0.045$& 0.039 \\
5s $^{2}S_{1/2}$ - 6p $^{2}P_{1/2}$ & $-0.062$       & $-0.123$  &   0.168          &  0.054   &  0.086          &  $-0.018$             &   $-0.005$           &    0.004             &$-0.009$              &$-0.002$&$-0.014$  \\
5s $^{2}S_{1/2}$ - 5p $^{2}P_{3/2}$ &  3.424         & $-0.176$  &   0.008          & $-0.211$ & $-0.289$        &   0.019               &   $-0.011$           &   $-0.001$           & 0.046                &$-0.062$& 0.048 \\
5s $^{2}S_{1/2}$ - 6p $^{2}P_{3/2}$ & $-0.175$       & $-0.163$  &   0.226          & $-0.011$ &  0.125          &  $-0.025$             &   $-0.006$           &    0.005             &$-0.014$              &$-0.0005$& 0.068 \\
&& \\
5p $^{2}P_{1/2}$ - 6s $^{2}S_{1/2}$ &  1.773         &  0.194    &  $-0.306$        &  0.007   &  0.018          &  $-0.0004$             &    0.007             &   $-0.011$           &$-0.0006$             &$-0.028$& 0.006  \\
5p $^{2}P_{1/2}$ - 5d $^{2}D_{3/2}$ &  4.011         & $-0.360$  &   0.085          & $-0.118$ & $-0.045$        &   0.010               &   $-0.008$           &   $-0.002$           & 0.027                &$-0.051$&$-0.074$  \\
&& \\
5p $^{2}P_{3/2}$ - 6s $^{2}S_{1/2}$ &  2.704         &  0.276    &  $-0.433$        &  0.003   &  0.020          &   0.002               &    0.009             &   $-0.016$           &$-0.0005$             &$-0.043$&$-0.004$   \\
5p $^{2}P_{3/2}$ - 5d $^{2}D_{3/2}$ &  1.867         & $-0.157$  &   0.036          & $-0.053$ & $-0.065$        &   0.005               &   $-0.004$           &   0.0005             & 0.013                &$-0.023$& 0.009 \\
5p $^{2}P_{3/2}$ - 5d $^{2}D_{5/2}$ &  5.581         & $-0.475$  &   0.109          & $-0.159$ & $-0.192$        &   0.015               &   $-0.010$           &    0.002             & 0.041                &$-0.072$& 0.028 \\	
		\end{tabular}
	\end{ruledtabular}
	\label{ContriE1RCCSD}
\end{table*}

In Table \ref{Abhfth}, we give the calculated values of $A_{hf}/g_I$ of the $^{113}$Cd$^+$ ion from the employed methods. Likewise the energies, the correlations trends
from the DHF to CCSD methods are found to be similar in this property. It has also been observed in the earlier studies that the signs of this quantity in the $D_{5/2}$
states are usually different at the DHF and CCSD methods in the alkali atoms \cite{sahoo1,sahoo2} and singly charged alkaline earth ions \cite{sahoo3} implying that electron correlation
effects are more than 100\% in these states. However, we do not find such trend in the considered ion. We also find the Breit interaction contributes insignificantly, but
the QED corrections to the $A_{hf}/g_I$ values in the $S$ states are seen to be quite large. Unlike other states, the correlation trends in the $F$ states are found to be
different. In this case, the final CCSD values are smaller than the DHF results. After multiplying with the respective $g_I$ values of the $^{111}$Cd$^+$ and
$^{113}$Cd$^+$ ions, we have given the $A_{hf}$ values of all the considered states in the same table. Since the mass differences between $^{111}$Cd and $^{113}$Cd are
very small, we have neglected the small changes in the wave functions to evaluate the $A_{hf}/g_I$ values for the $^{111}$Cd$^+$ ion here. We also compare our estimated
$A_{hf}$ values of these ions with the available experimental values and other calculations. Very precise values of these quantities for the ground state of
both the isotopes are available \cite{CdIIhysexp1996,CdIIhysexp2006,CdIIhysexp2012,CdIIhysexp2013,CdIIhysexp2015}. Among them the most precise values are reported
 for the  $^{111}$Cd$^+$ and $^{113}$Cd$^+$ ions as 14530.5073499(11) MHz \cite{CdIIhysexp2012} and 15199.8628550192(10) MHz \cite{CdIIhysexp2015}, respectively. We have only quoted
these values up to three decimal places in Table \ref{Abhfth}. Though measurements of $A_{hf}$ values are not reported precisely in the other states, however a preliminary
measurement of hyperfine splitting in the $5p \ ^2P_{3/2}$ state of the $^{113}$Cd$^+$ ion suggests that its $A_{hf}$ value is about 400 MHz \cite{CdIIhysexp1996}. It can be seen that our
calculations agree with the available experimental values quite well. Therefore, we also anticipate that the values estimated for the other states will have similar
accuracies. We had also studied these properties before using the CCSD method \cite{Dixithys2008} without accounting for the Breit and QED interactions. Moreover, we had only
included few non-linear terms from the non-terminating series of $e^{T^{\dagger}} O e^T$ and $e^{T^{\dagger}} e^T$ in Eq. (\ref{preq}). Later, we have developed procedures
to include contributions from these series self-consistently to infinity number (see e.g. \cite{sahoo2,sahoo4}). These are the reasons why we obtain improved values of
$A_{hf}$ in this work. Also, details of correlation trends are analyzed by us here.

To gain better understanding of roles of electron correlation effects in the evaluation of the $A_{hf}/g_I$ values in Cd$^+$, we present individual contributions from
various CCSD terms in Table \ref{ContriAhfRCCSD}. We give contributions as $\overline{O}$, corresponding to effective one-body part of $e^{T^{\dagger}} O e^T$, and other
terms linear in $T$, $S_v$ and their hermitian conjugate ($h.c.$) operators. Corrections due to normalization of the wave functions are also given explicitly. Contributions
from the remaining non-linear terms are given together. The difference between the DHF value and the $\overline{O}$ contribution for a given state implies the role of core
correlations. It can be seen that in the states with angular momentum $J=1/2$, the core correlations decrease the values from DHF method while for other angular momentum states
it is increasing. As we had discussed in our earlier work \cite{sahoo5}, the terms $OS_{1v}$ and $OS_{2v}$ (along with their $h.c.$ terms)
represent for the all-order pair-correlation and core-polarization effects, respectively. Other terms can be interpreted as the higher order correlations representing one
of these kinds. It can be observed from the above table that the pair-correlation and core-polarization effects are equally important in the determination of $A_{hf}/g_I$
values in the $S$ states; the former types are more significant in the $5S$ and $6S$ states while the later effects are slightly larger than the former in the $7S$ state.
In the $P$ states also, the pair-correlation effects play the dominant roles. However, the core-polarization effects are found to play significant roles in the states
with higher angular momentum. Contributions from the normalization of the wave functions and the non-linear terms appearing through property evaluation expressions are
also found to be non-negligible.

\begin{table}[t]
	\caption{Correlation trends in the determination of $\alpha^{E1}$ values (in a.u.) using various approximations in Cd$^+$.}
	\begin{ruledtabular}
		\begin{tabular}{lcccc}
		Method   &   $5s \ ^{2}S_{1/2}$   & $5p \ ^{2}P_{1/2}$  & \multicolumn{2}{c}{$5p \ ^{2}P_{3/2}$} \\
		\cline{4-5}\\
		            &  Scalar             &   Scalar            & Scalar  &  Tensor \\
			\hline
            & & \\
            DHF &  36.792  &  33.913 & 39.927 &  $-6.433$ \\
            MBPT(3) & 22.273 & 27.964 & 31.982 & $-4.327$ \\
            CCSD &  24.637   & 25.817  &  30.594 & $-3.681$  \\
            CCSD$+$Experiment & 25.2(6) & 25.2(1.5) & 28.1(6) & $-2.3(4)$ \\
		\end{tabular}
	\end{ruledtabular}
	\label{Polzcorr}
\end{table}

\begin{table}[t]
	\caption{Breakdown of contributions to $\alpha^{E1}$ values (in a.u.) of the ground, $5p \ ^2P_{1/2}$ and $5p \ ^2P_{3/2}$ states in Cd$^+$ by combining experimental
	data and calculations from the CCSD method. Intermediate contributions to ``Main'' are quoted, in which precisely available experimental E1 matrix elements are used.
	We also use the E1 matrix elements from our previous work Ref. \cite{YuBBR2017}, that are not discussed in this paper, to estimate their contributions. Both the ``Tail'' and
	``Core-valence'' contributions are given from the DHF method, while the ``Core'' correlations are determined using RPA.}
	\begin{ruledtabular}
		\begin{tabular}{lccc}
			Contributions          &  E1 amplitude   & Source                    &  $\alpha^{E1}$   \\
			\hline
			& & \\
            \multicolumn{4}{c}{Scalar polarizability of the ground state}\\
            From Main: \\
	    $5p \ ^{2}P_{1/2}$                   &  1.910(4) & Experiment \cite{CdII5plifetime2006}   &  6.047(25)  \\
            $(6-12)p \ ^{2}P_{1/2}$             &  & This work  \cite{YuBBR2017}       &  0.016       \\
            $5p \ ^{2}P_{3/2}$                  &  2.713(5) & Experiment \cite{CdII5plifetime2006}   & 11.551(43)       \\
            $(6-12)p \ ^{2}P_{3/2}$           &  & This work  \cite{YuBBR2017}        &  0.011       \\
            \multicolumn{2}{l}{From $4d^9 \ 5s5p$
            configurations}                      &  This work \cite{YuBBR2017}         &  2.6(5)        \\
            Tail                                                  &                      &  This work              &  0.012              \\
            Core-valence                           &             &                      This work           & $-0.018$               \\
            Core                                                &                     &    This work             &  4.986               \\
            \hline
            & & \\
            \multicolumn{4}{c}{Scalar polarizability of the $5p \ ^2P_{1/2}$ state}\\
            From Main:\\
            $5s \ ^{2}S_{1/2}$                   &  1.910(4) & Experiment \cite{CdII5plifetime2006}   & $-6.047(25)$ \\
            $6s \ ^{2}S_{1/2}$                    &  1.72(12) & Experiment \cite{CdI-IIexpt2004}       &  5.57(80)    \\
            $(7-12)s \ ^{2}S_{1/2}$              &    & This work  \cite{YuBBR2017}                       &  0.393     \\
            $5d \ ^{2}D_{3/2}$                    &  3.08(13) & Experiment \cite{CdI-IIexpt2004}       &  15.2(1.3) \\
            $6d \ ^{2}D_{3/2}$                    &   &  This work  \cite{YuBBR2017}          &   3.546    \\
            $(7-12)d \ ^{2}D_{3/2}$               &    &  This work \cite{YuBBR2017}            &   0.809    \\
            $4d \ ^{9} 5s\ ^{2}~^{2}D_{3/2}$      &   0.49(2) & Experiment \cite{CdI-IIexpt2004}       &  0.57(2)   \\
            Tail                                             &        &    This work                          &  0.375              \\
            Core-valence                                   &           &   This work                        & $-0.193$               \\
            Core                                              &         &  This work                           &  4.986               \\
            \hline
            & & \\
            \multicolumn{4}{c}{Scalar polarizability of the $5p \ ^2P_{3/2}$ state}\\
            From Main:\\
            $5s \ ^{2}S_{1/2}$                 &  2.713(5) & Experiment \cite{CdII5plifetime2006}   & $-5.775(21)$ \\
            $6s \ ^{2}S_{1/2}$                   &  2.31(9) & Experiment \cite{CdI-IIexpt2004}        &  5.37(42) \\
            $(7-12)s \ ^{2}S_{1/2}$             &    & This work   \cite{YuBBR2017}                    &  0.426    \\
            $5d \ ^{2}D_{3/2}$                 &   1.57(6) & Experiment \cite{CdI-IIexpt2004}       &  2.09(16) \\
            $(6-12)d \ ^{2}D_{3/2}$             &     &  This work   \cite{YuBBR2017}                   &  0.209    \\
            $5d \ ^{2}D_{5/2}$                 &  4.62(5) & Experiment \cite{CdI-IIexpt2004}        & 18.06(39) \\
            $(6-12)d \ ^{2}D_{3/2}$           &    &   This work  \cite{YuBBR2017}                     &  1.961    \\
            $4d^95s^2 \ ^{2}D_{3/2}$ &         0.29(4) & Experiment \cite{CdI-IIexpt2004}       &  0.11(3)   \\
            $4d^95s^2 \ ^{2}D_{5/2}$ &        0.58(4) & Experiment \cite{CdI-IIexpt2004}       &  0.54(8)   \\
            Tail                                      &       &This work                               &  0.341              \\
            Core-valence                             &          & This work                            & $-0.189$               \\
            Core                                        &       & This work                              &  4.986               \\
            \hline
            & & \\
            \multicolumn{4}{c}{Tensor polarizability of the $5p \ ^2P_{3/2}$ state}\\
            From Main:\\
            $5s \ ^{2}S_{1/2}$                &  2.713(5) & Experiment \cite{CdII5plifetime2006}   &  5.775(21) \\
            $6s \ ^{2}S_{1/2}$                 &  2.31(9) & Experiment \cite{CdI-IIexpt2004}        & $-5.37(42)$  \\
            $(7-12)s \ ^{2}S_{1/2}$           &    & This work  \cite{YuBBR2017}                      & $-0.426$     \\
            $5d \ ^{2}D_{3/2}$                 &  1.57(6) & Experiment \cite{CdI-IIexpt2004}        &  1.67(13)  \\
            $(6-12)d \ ^{2}D_{3/2}$          &    & This work  \cite{YuBBR2017}                     &  0.167     \\
            $5d \ ^{2}D_{5/2}$                  &  4.62(5) & Experiment \cite{CdI-IIexpt2004}        & $-3.61(8)$   \\
            $(6-12)d \ ^{2}D_{3/2}$          &    & This work  \cite{YuBBR2017}                     & $-0.392$     \\
            $4d^95s^2 \ ^{2}D_{3/2}$                &   0.29(4) & Experiment \cite{CdI-IIexpt2004}       &  0.09(3)   \\
            $4d^95s^2 \ ^{2}D_{5/2}$                &   0.58(4) & Experiment \cite{CdI-IIexpt2004}       & $-0.11(2)$   \\
		    Tail                             &        &   This work                            & $-0.084$    \\
            Core-valence                                &      & This work                                &  0.010 \\
		\end{tabular}
	\end{ruledtabular}
	\label{polari}
\end{table}

In Table \ref{E1-MEs}, we present the magnitudes of the reduced E1 matrix elements for some of the important transitions of Cd$^+$. As can be seen, in most of the cases the
DHF values are very large and the correlation effects reduce the magnitudes. This trend is different from the studies of the magnetic dipole hyperfine structure constants.
We also find that the values are reduced slightly from the DHF values in the MBPT(2) method, but then they are reduces drastically in the LCCSD method. The CCSD method gives values
intermediate between the MBPT(2) and LCCSD methods, suggesting that the non-linear terms of the RCC method cancels with some of the correlation effects arising through the
LCCSD method. It is also noticed that the DHF values in the 5s $^{2}S_{1/2}$ - 6p $^{2}P_{1/2}$ and 5s $^{2}S_{1/2}$ - 6p $^{2}P_{3/2}$ transitions have opposite signs than
the results after including the correlation effects. It means that the electron correlation effects are larger than 100\% in these transitions. We also quote uncertainties to the
CCSD results by analyzing contributions due to the partial triples contributions. We have compared our calculations with the values that are inferred from the lifetime
measurements of many excited states reported in the literature \cite{CdII5plifetime2006,CdI-IIexpt2004}. Unlike the hyperfine structure constants, we find large discrepancies
between our results with the experimental values in this property. We also give E1 matrix elements from other calculations that are reported using the DHF method, MBPT(2)
method and RCC theory similar to our LCCSD approximation (defined as all-order singles and doubles (SD) method) \cite{UISafro2011}. In our work, we use Gaussian type of
orbitals (GTOs) to define the single particle matrix elements, however B-spline polynomials were used in Ref. \cite{UISafro2011}. Nevertheless, we find good agreement
between these two works in the DHF and MBPT(2) methods. Large differences are observed among the values from the LCCSD and SD methods. Since the non-linear terms through the
CCSD method increase values from the LCCSD method, we also anticipate that values from the SD method will be larger after taking into account the non-linear contributions.
Therefore, large discrepancies between the CCSD values and the experimental results may be able to address by including full triples and quadruple excitations in the
RCC theory.

 We also give contributions from the individual terms of the CCSD method to the E1 amplitudes in Table \ref{ContriE1RCCSD}. It can be found from the differences
between the DHF values and the $\overline{O}$ contributions that the core correlations are negligible in this property. The pair-correlation effects arising through the
$S$ states in the $S-P$ transitions and the $P$ states in the $P-D$ transitions are found to be quite large. However, this trend seems to be reverse for the core-polarization
correlation contributions. Correlation contributions from the other terms are found to be very small. This indicates that it would be necessary to include correlation
effects that can influence the pair-correlation and core-polarization correlations strongly in order to achieve more precise values of the E1 matrix elements.

 We intend to present now the dipole polarizabilities of the atomic states of Cd$^+$, which are not yet investigated in the literature. These quantities are very useful
for the high-precision measurements. Thus, we would like to estimate them more precisely for the general interest. It to be noted that the allowed transitions among 
low-lying states are very useful for the cooling mechanism of the singly charged ions. Knowledge of accurate values of $\alpha^{E1}$ for the states associated
with these transitions will be required for such studies. In this view, we determine the $\alpha^{E1}$ values of the ground, $5p \ ^2P_{1/2}$ and $5p \ ^2P_{3/2}$
states of Cd$^+$ here. 

First, we present {\it ab initio} results of $\alpha^{E1}$ in Table \ref{Polzcorr} from the DHF, MBPT(3) and CCSD methods in the perturbative and finite-field approaches as 
described in Sec. \ref{sec3}. As can be seen, the DHF method predicts relatively larger values and the MBPT(3) method, which uses energies and E1 matrix elements at the 
level of MBPT(2) method approximation, brings down the results. Then, the CCSD method gives intermediate value between those two lower-order methods for the ground state 
while it decreases further the $\alpha^{E1}$ values in all other states. Since the correlation effects with respect to the DHF values are found to be strong in these 
quantities, we would like reduce their uncertainties by substituting the precise data of E1 matrix elements from the lifetime measurements of the low-lying states of Cd$^+$ 
and experimental energies with the CCSD results. For this purpose, we have used the sum-over-states approach to estimate the most accurate data and quote the values 
in the above table as ``CCSD$+$Experiment''. As seen, these semi-empirical values are differing significantly from the corresponding CCSD results. The reason for this 
can be obvious from the comparison of the E1 matrix elements from the CCSD method with the experimental values in Table \ref{E1-MEs}. We have also quoted the 
uncertainties to the results obtained from the combined experimental and CCSD results. Break down of various contributions to these quantities for the CCSD$+$Experiment
approach are given in Table \ref{polari}. It can be noticed that we have also used many matrix elements in this approach from our CCSD method that are not quoted in Table
\ref{E1-MEs} but that are given in our another recent work \cite{YuBBR2017}. We also give contributions from ``Tail'', core-valence and core correlation
contributions to the sum-over-states approach in Table \ref{polari}. Since the Tail and core-valence correlation contributions are extremely small, they are estimated using
the DHF method. Comparatively, the core correlation contributions to the scalar polarizabilities are larger. We have estimated these contributions using random phase 
approximation (RPA) as described by us earlier \cite{yashpal}. Our final values for the ground and $5p \ ^2P_{3/2}$ states are found to be quite precise, but we still get 
quite sizable amount of uncertainty to the $\alpha^{E1}$ value of the $5p \ ^2P_{1/2}$ state owing to large uncertainty associated with the 
$5p \ ^2P_{1/2} - 6s \ ^{2}S_{1/2}$ transition.

\section{Summary} \label{sec4}

We have investigated electron correlation trends in the energies, hyperfine structure constants and electric dipole matrix elements of the singly charged ion. We have
employed mean-field, finite-order perturbation and all-order coupled-cluster theories in the relativistic framework to carry out these analyses. Our results employing
singles and doubles approximated relativistic coupled-cluster method are found to be agreeing very well with the experimental results, but the electric dipole matrix elements
do not agree well with those are extracted out from the lifetime measurements. Further theoretical studies are required to explain reasons for such large discrepancies. We
have also given hyperfine structure constants of many states in which experimental values are not known. These results will be useful to guide the experimentalists to
measure them precisely and test the validity of our calculations. Correlation trends to the {\it ab initio} values of the dipole polarizabilities in the first three low-lying
states of Cd$^+$ are also given using the Dirac-Hartree-Fock approximation, third-order many-body theory and using the singles and doubles approximated coupled-cluster
method. We also deduce them more accurately by replacing the calculated E1 matrix elements by the precisely known electric dipole matrix elements wherever available and 
combining with the experimental energies in a sum-over-state approach. These quantities will be helpful to carry out high-precision measurements in the Cd$^+$ ion.

\section*{Acknowledgement}

C.-B. L. acknowledges support from National Science Foundation of China (Grant No. 91536102 and 91336211), the Strategic Priority Research Program
of CAS (Grant No. XDB21030300) and the National Key Research and Development Program of China (Grant No. 2017YFA0304402). B. K. S. acknowledges financial support from CAS through the PIFI fellowship under the project number 2017VMB0023.
Computations were carried out using Vikram-100 HPC cluster of Physical Research Laboratory (PRL), Ahmedabad, India. Y.Y. acknowledges support from NSFC
(Grant No. 91536106).

\end{document}